\documentclass[11pt]{pazh}

\usepackage{psfig}

\usepackage[T2A]{fontenc}
\usepackage[utf8]{inputenc}

\usepackage{latexsym}
\usepackage{amssymb}

\begin{document}

\setcounter{page}{1}

\sloppypar

\title{\bf On the characteristics of tidal structures of interacting galaxies}

\author{Y.H. Mohamed\inst{1,2}, V.P. Reshetnikov\inst{1}, N.Ya. Sotnikova\inst{1}} 

\institute{St.Petersburg State University, Universitetskii pr. 28, Petrodvoretz, 
198504 Russia
\and
Astronomy Department, National Research Insitute of Astronomy and Geophysics,
Cairo 11421, Egypt
}

\titlerunning{Tidal structures}

\abstract{We present the results of our analysis of the geometrical tidal tail 
characteristics for nearby and distant interacting galaxies. The sample includes 
more than two hundred nearby galaxies and about seven hundred distant ones. The 
distant galaxies have been selected in several deep fields of the Hubble Space 
Telescope (HDF-N, HDF-S, HUDF, GOODS, GEMS) and they are at mean redshift 
$\langle z \rangle = 0.65$. We analyze the distributions of lengths and thicknesses 
for the tidal structures and show that the tails in distant galaxies are shorter 
than those in nearby ones. This effect can be partly attributed to observational
selection effects, but, on the other hand, it may result from the general evolution 
of the sizes of spiral galaxies wih $z$. The location of interacting galaxies on the 
galaxy luminosity ($L$) -- tidal tail length ($l$) plane are shown to be explained 
by a simple geometrical model, with the upper envelope of the observed distribution 
being $l \propto \sqrt{L}$. We have solved the problem on the relationship between 
the observed distribution of tail flattening and the tail length in angular measure 
by assuming the tidal tails to be arcs of circumferences visible at arbitrary angles 
to the line of sight. We conclude that the angular length of the tidal tails visually
distinguished in nearby and distant galaxies, on average, exceeds 180$^{\rm o}$.
}
\titlerunning{Tidal structures}
\maketitle

\section{Introduction}

Tidal structures (the tails and the bridges connecting galaxies) are transient
features emerging during close encounters and mergers of galaxies (Toomre and
Toomre 1972). As a rule, tidal structures have a low optical surface brightness
($\mu(B) \approx 24^m - 25^m/\Box''$) and they are observed in several percent
of galaxies in the local Universe (for an overview, see
Sotnikova and Reshetnikov 1998a; Reshetnikov and Sotnikova 2001).

Tidal structures are interesting for many reasons.
For example, dwarf galaxies can be formed from their
matter (Duc 2011). The formation of massive clumps,
up to $10^{8}$~M$_{\odot}$, by the gravitational collapse
of stars and gas clouds extended into the tidal
tail are commonly observed in numerical simulations
of galaxy interactions (see, e.g., Barnes and Hernquist
1992; Elmegreen et al. 1993). Such clumps 
can give rise to dwarf galaxies.

The current star formation rate in tidal tails is
occasionally high; the star-forming regions (HII regions)
can be arranged in the tail uniformly, as, for
example, in the tail of NGC~4676\,A (Sotnikova and
Reshetnikov 1998b). In the discs of normal galaxies,
star formation usually takes place in giant H$_2$
complexes. Tidal structures are formed from the diffuse gas
stretched from the outermost regions of a galaxy. The
question about the star formation mechanism in these
structures is open. For example, such a mechanism
may be associated with global gravitational instability
in tails (Sotnikova and Reshetnikov 1998b).

The morphology of tails and bridges is determined
by the global dynamical structure of galaxies. For
example, it has emerged that the length of tidal tails
depends not only on the impact parameters and the
relative velocity of interacting galaxies but also on the
mass distribution of dark matter (see, e.g., Dubinski
et al. 1996, 1999; Mihos et al. 1998; Springel and
White 1999). Having analyzed the results of numerical
simulations, Dubinski et al. (1996) noticed that
if the mass of dark matter is large enough, then the
forming tidal tails turn out to be too short and indistinct.
At the same time, in interacting and merging
systems, we often observe very extended tidal structures
stretching to distances as large as 50--100 kpc.

The conclusion that extended tidal tails can serve
as an indicator for the presence of dark matter in
the outermost regions of galaxies, where the
HI disc is invisible, is also corroborated by
kinematic data. Using the primary component of
the famous interacting system NGC~4676 (the Mice)
as an example, Sotnikova and Reshetnikov (1998b)
concluded that the kinematics of the tidal tail (its
length exceeds 40 kpc) is consistent only with the
model of a dark halo whose mass within the optical
tail exceeds the total mass of the galactic disc and
bulge by several times.

Mutual agreement between the kinematic and
morphological analyses suggests that conclusions
about dark halo characteristics can be made even though
the spectroscopic data are not available and we have
only photometric data (in the case of distant galaxies
with tidal features having a low surface brightness).

The occurrence of tidal structures at different redshifts
is of great interest, because it reflects the
change in the rate of galaxy mergers and interactions
with time (Reshetnikov 2000; Bridge et al. 2010).

The main goal of this paper is to study basic 
geometrical characteristics of the tidal tails in several
hundred nearby and distant interacting galaxies so as
to reach conclusions about the relationship between
these characteristics and the global parameters of the
galaxies themselves. All numerical values in the paper
are given for the cosmological model with a Hubble
constant of 70 km s$^{-1}$ Mpc$^{-1}$ and 
$\Omega_m=0.3$, $\Omega_{\Lambda}=0.7$.

\section{The samples of galaxies and the measured parameters}

\subsection{Nearby and distant galaxies with tidal structures}

To study the characteristics of nearby galaxies,
we considered two samples: (1) KPG -- binary galaxies
from the catalog by Karachentsev (1987); (2) SDSS --
galaxies with tidal tails from the catalog by
Nair and Abraham (2010) based on the visual classification
of 14\,034 objects from the Sloan Digital
Sky Survey. The first sample includes 44 galaxies
for which the characteristics of their tidal structures
could be measured in the Digital Sky Survey (DSS);
the second sample includes 182 objects with tidal
structures clearly distinguishable in the SDSS. Since
many of the galaxies studied exhibit not one but two
tails, the number of tidal tails we measured (64 and
266 in the KPG and SDSS samples, respectively)
exceeds the number of galaxies.

When studying distant galaxies, we used the
catalog of interacting galaxies found in deep fields
of the Hubble Space Telescope (Mohamed and
Reshetnikov 2011). This catalog includes data for
about seven hundred candidates for galaxies with
tidal structures ($z \leq 1.5$) in several Hubble deep fields (HDF-N,
HDF-S, HUDF, GOODS, GEMS). The investigated
fields partly overlap (e.g., HUDF is part
of GOODS) and, in these cases, the characteristics
of galaxies were estimated from the deeper field. We
studied a total of 867 tidal structures in all deep fields.

\subsection{Parameters of tidal structures}

\begin{figure}
\centerline{\psfig{file=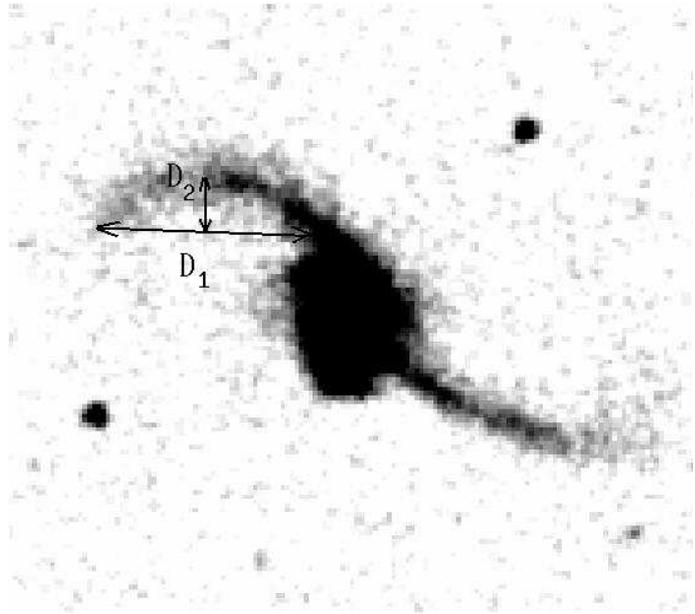,width=9cm,clip=}}
\caption{Determining $D_1$ and $D_2$ for tidal tails using a
high-contrast DSS image of the nearby merging system
NGC~2623 as an example: $D_1$ is the lenght of the segment connecting
the beginning and the end of the tidal structure, $D_2$ is the size of the
perpendicular drawn from the center of this segment to the tail.}
\end{figure} 

For all interacting galaxies, we determined the
geometrical characteristics of their tidal structures
from high-contrast optical images: the length of the
tail from the edge of the galactic disc to its end ($l$)
measured along the curved tail,
the half-length width ($h$), and the ratio $k = D_2/D_1$,
which is a measure of the curvature (see Fig.~1).
It is obvious that $k=0$ for a straight tail and $k>0$
for a curved one. If the tail is assumed to be an arc of a
circumference, then $k$ for an arc of 180$^{\rm o}$ is equal to 0.5.
(All the measurement were done within the 
surface brightness limit of $\mu(I)\approx 25-26$.)

When analyzing the objects from the KPG sample,
we used blue DSS images; for the objects from
Nair and Abraham (2010), we took data in the $g$
band ($\lambda_{eff}=4686$\,\AA). The characteristics of galaxies
from the deep fields were determined in the F814W
(HDF-N, HDF-S), F775W (HUDF), and F850LP
(GOODS and GEMS) filters (at redshift $z\sim 1$)
these filters roughly correspond to the $B$ band in the
reference frame associated with the galaxy).
To verify the nature of the tidal structures in
faint distant galaxies, we performed
their aperture photometry at several points along the
structures. The mean surface brightness of
the tails recalculated to the $B$ band in the
reference frame associated with the galaxy
($\langle \mu_B \rangle = 24.7\pm1.1(\sigma)$)
turned out to be close to its typical values
for the tidal structures of local galaxies (see, e.g.,
Schombert et al. 1990; Reshetnikov 1998).

The redshift estimates and apparent magnitudes
are available for all of the distant galaxies we considered
(Fernandez-Soto et al. 1999; Williams et al. 2000;
Sawicki and Mallen-Ornelas 2003; Wolf et al. 2004;
Glazebrook et al. 2006; Coe et al. 2006; Balestra
et al. 2010), allowing to find the galaxy luminosities.
The mean redshift of our sample of distant
interacting galaxies is 
$\langle z \rangle = 0.65 \pm 0.31(\sigma)$.

\section{Results and discussion}

\subsection{Geometrical characteristics of tidal tails}

\begin{figure}
\centerline{\psfig{file=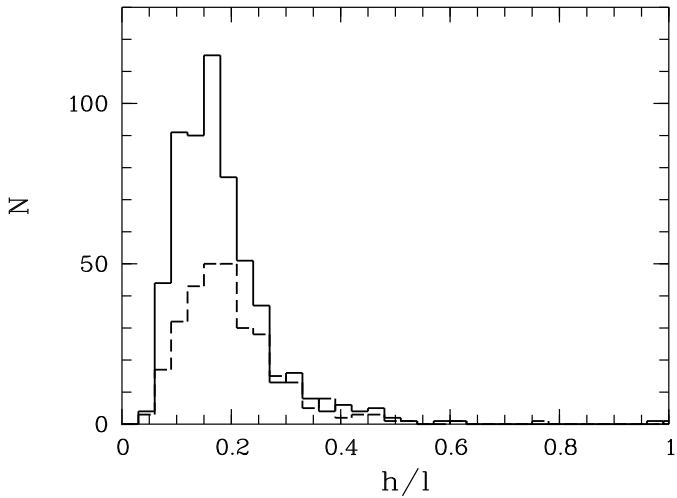,width=9cm,angle=-90,clip=}}
\caption{The distributions of relative tail thicknesses ($h/l$) for nearby 
(dashed line) and distant (solid line) galaxies with $l \geq 10$ kpc.}
\end{figure} 

Figure~2 shows the observed distributions of
thicknesses for the tidal structures in nearby and
distant galaxies with relatively long ($l \geq 10$ kpc)
tails. Both distributions are similar and they demonstrate
peaks of observed flattening at $h/l \approx 0.15$. As we
see from the figure, there are very thin ($h/l \leq 0.1$) and
relatively wide ($h/l \geq 0.2$) tails.

\begin{figure*}
\centerline{\psfig{file=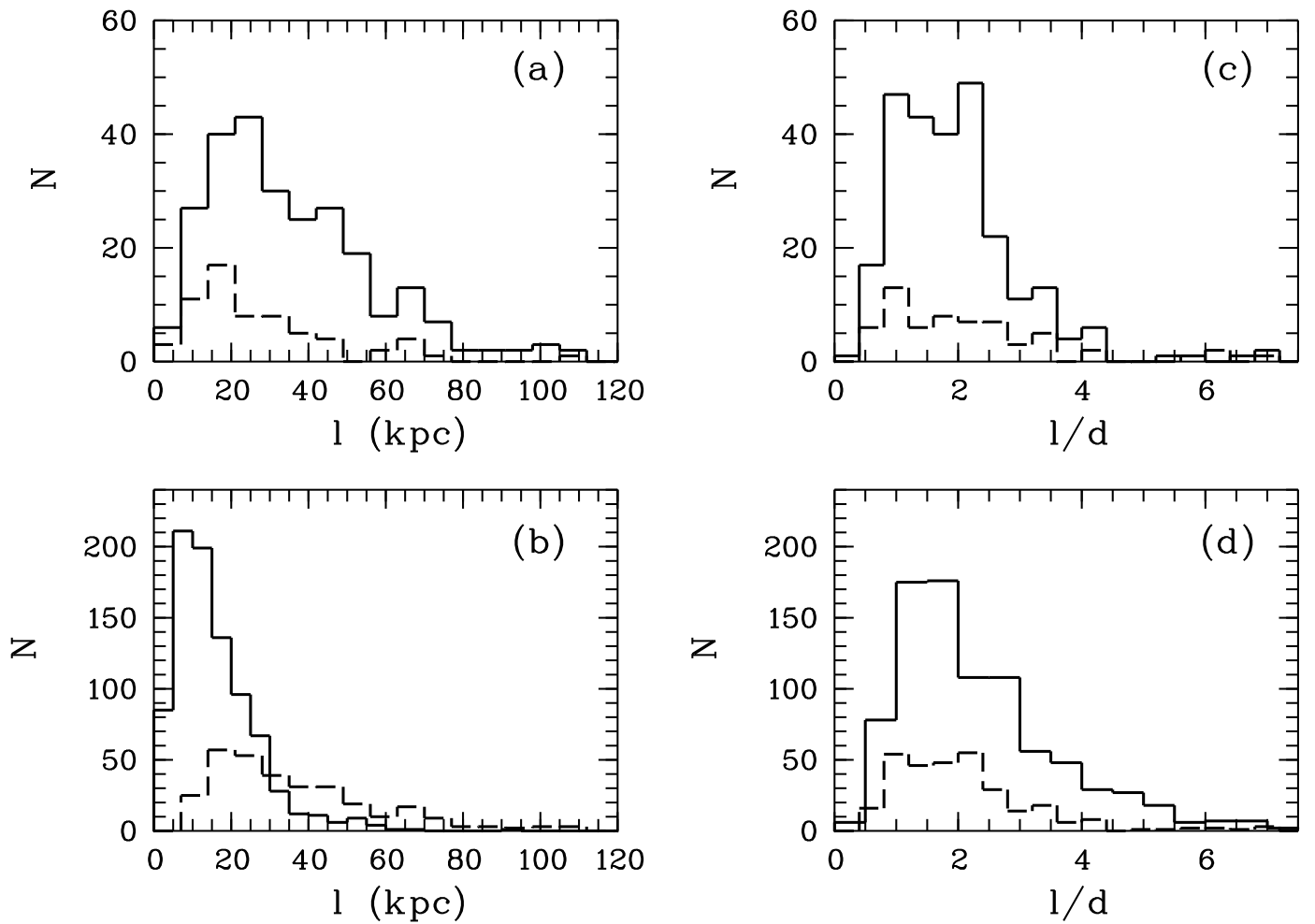,width=22.5cm,angle=-90,clip=}}
\caption{The distributions of absolute (a, b) and relative (c, d) tidal tail
lengths for nearby and distant galaxies: (a) the solid and
dashed lines indicate the distributions of tail lengths for the nearby galaxies
from the SDSS and KPG samples, respectively; (b) the distant and nearby (the
combined KPG + SDSS sample) objects are indicated by the solid and dashed lines,
respectively; (c) the relative tail lengths ($l/d$) in the SDSS (solid line)
and KPG (dashed line) samples; (d) the distributions of distant (solid
line) and nearby (KPG + SDSS) galaxies in $l/d$. }
\end{figure*}

Figure~3a shows the distributions of tail lengths
in the KPG and SDSS samples of nearby galaxies.
The distributions are similar and they exhibit global
maxima at $l \approx 20$ kpc. Figure 3b displays the analogous
distribution for distant galaxies in comparison
with the combined (KPG + SDSS) distribution for
nearby ones. In this figure, we see the difference
between the two distributions: for distant galaxies,
the maximum occurs at $l \approx 10$ kpc. Moreover,
among the most distant galaxies, extended ($l > 30$ kpc)
tidal structures are less common. It is
somewhat premature to conclude about
the evolution of the tail lengths with $z$,
because the statistics of the lengths of tidal structures
for distant galaxies is undoubtedly distorted by observational
selection. The most important selection
effects, that lead a rapid decrease in brightness, are
the cosmological surface brightness
dimming (Tolman's effect) and the $k$ correction.
When measuring the tail lengths within a
fixed isophote, we will obtain systematically progressively
shorter structures as the redshift increases.

To avoid these effects, at least partially, we measured
the tail length in diameters of the galaxy's main body ($d$).
The diameter was estimated simultaneously with the tail
length and it corresponds to approximately the same
brightness level. Since the cosmological dimming and the
$k$ correction must
reduce the measured angular size of a galaxy, the ratio
of the tail length to the diameter of the main galaxy
is affected by these effects to a lesser extent. Figure~3c
shows the observed distributions of relative tail
lengths for the samples of nearby galaxies. These distributions
are similar and they exhibit peaks at $l/d \approx 1.5-2$.
Figure 3d displays the distribution of relative
tail lengths for distant galaxies in comparison with
the combined distribution for nearby ones. We see
from this figure that there is no statistically significant
difference between the two distributions: the relative
length of the tidal tails in distant galaxies, just as in
nearby ones, exhibits a maximum at $l/d \approx 1.5-2$
(the corresponding median values are 1.99 and 1.80).

The relationships between the relative ($l/d$) and
linear tail lengths for nearby and distant galaxies
shown in Fig.~4 differ noticeably. Previously,
Elmegreen et al. (2007) considered the $l/d - l$ plane for
about twenty distant (from the GEMS and GOODS
fields) and nearby galaxies with extended tidal structures.
They found the tails of nearby galaxies to be a
factor of 2.7 longer than those of distant ones. Our
data confirm this result: as we see from Fig.~4, the
tails of nearby galaxies at fixed $l/d$ are appreciably
(on average, by a factor of 2.6) longer. Such a good
quantitative agreement between the results obtained
from samples differing by several tens of times is
remarkable, although it may be partly accidental.

\begin{figure}
\centerline{\psfig{file=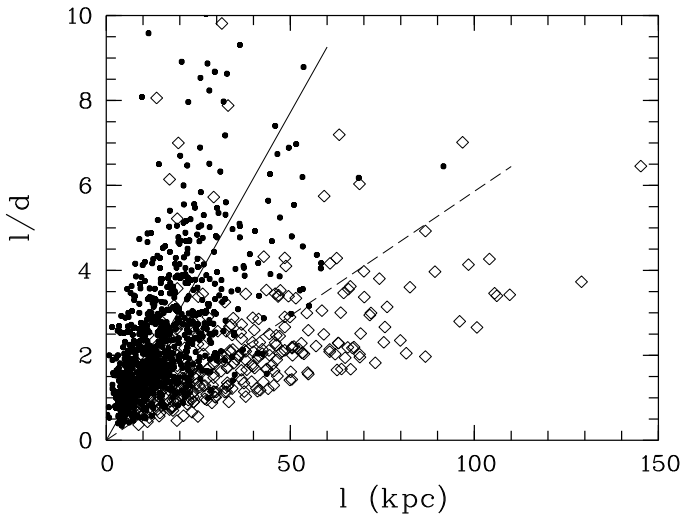,width=9cm,angle=-90,clip=}}
\caption{Nearby (diamonds) and distant (filled circles) galaxies on the
$l - l/d$ plane. The linear regression is indicated by the solid
and dashed straight lines for distant and nearby galaxies, respectively.
In order to reconcile the slopes of these dependencies, 
tail lengths for distant galaxies should be increased by a factor of 2.6.}
\end{figure} 

The observed difference between the linear lengths
of tidal structures can be attributed, at least partly,
to observational selection. On the other hand, it can
reflect the actually observed evolution of the sizes of
spiral galaxies and, as a consequence, the sizes of
their tidal structures (for a discussion, see the Conclusions).

\subsection{The galaxy luminosity -- tail length relationship}

The relationship between the absolute magnitude
of a galaxy and the length of its tidal tail is shown
in Fig.~5. This relationship can indirectly reflect 
the connection between the lenght of tails and 
the global dynamical structure of galaxies expected 
from theoretical considerations (Dubinski et al. 1996, 1999; Mihos
et al. 1998; Springel and White 1999).

\begin{figure}
\centerline{\psfig{file=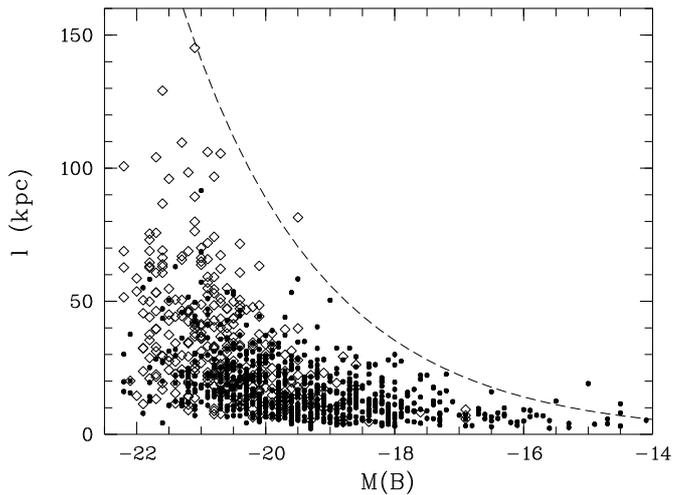,width=9cm,angle=-90,clip=}}
\caption{Total galaxy luminosity versus tail length. The data for nearby 
and distant galaxies are indicated by the diamonds and
filled circles, respectively. The dashed line indicates the dependence 
$l \propto \sqrt{L}$ following from simple geometrical considerations
(see the text).}
\end{figure} 

The formation of a tidal tail is a purely energetic effect.
The disc stars must acquire the energy sufficient to escape
out to a distance equal to the tail length. The energy
acquisition depends on two competing factors:
the galaxy interaction time and the strenght of perturbation.
Both depend on the mass distribution, primarily in
the dark halo, or on the potential gradient. Bright
galaxies probably possess, on average, massive dark
halos. In this case, the action of the perturbing force
will be more likely impulsive than resonant. This
implies that such galaxies will form less extended tails
during their interaction.

Another difficulty arises in the case of low-mass
dark halos. Although the tidal tails turn out to be very
long, the interacting galaxies merge very
rapidly. That is why the relation between the dark halo
mass and the tidal tail length is ambiguous. It should
also be noted that the lenght of the tidal tail depends
strongly not only on the depth of the potential well but
also on the initial distribution of visible matter (stars
and gas) in the disk of the parent galaxy (Sotnikova
and Reshetnikov 1998b). If, for example, the tidal
tails are assumed to be formed mainly from gas in
which a starburst subsequently occurs, then extended
features can be obtained in models with large dark
halo mass.

Thus, it is rather difficult to construct an unambiguous
relation between the tidal tail length and the
dark halo mass that can be associated with the total
galaxy luminosity. It is much easier to try to find
the direct relationship between the galaxy luminosity
and the tidal tail length suggesting a simple model.

Let us consider the following geometrical model. We
suppose that a fixed fraction of the galactic mass goes
into the galaxy's tidal tail and that this matter has
the same mass-to-light ratio as that of the galaxy
itself. The surface brightness of the tail will then be
determined by its luminosity divided by the area that
in our notation is equal to $\beta \times l^2$, where $\beta = h/l$.
At fixed surface brightness and relative width of the tail,
we can then immediately obtain the relation between
the total galaxy luminosity and the tail length. In the
$B$ band, this relation is
\begin{equation}
{\rm lg}\,l = -0.2\,(M_\mathrm{tot} + \alpha) - 0.5\,{\rm lg}\,\beta +
0.2\,\mu_\mathrm{tail} - 7.31,
\end{equation}
where the tail length $l$ is in kpc, $M_\mathrm{tot}$ is the absolute
magnitude of the galaxy, $\alpha$ is the difference
between the absolute magnitude of the tail and the galaxy
(for example, if the tail luminosity is 10\%
of the total galaxy luminosity, then $\alpha=2.^m5$),
$\beta$ is the relative width of the tail, and $\mu_\mathrm{tail}$ is
its surface brightness.

The dashed line in Fig. 5 indicates dependence (1)
(it is $l \propto \sqrt{L}$, where $L$ is the total luminosity of the
galaxy) for the following parameters:
$\alpha = 1.^m75$, i.e. the luminosity of the tidal structure
is 20\% of the total luminosity, 
$\beta=0.15$ and $\mu_\mathrm{tail}=26^m/\Box\arcsec$.
Wider, brighter, and relatively less luminous tails
must lie under this model dependence. As we see
from the figure, the simple geometrical model describes
satisfactorily the upper envelope of the actual
distribution of galaxies. This implies that the visually
distinguished optical tidal tails constitute a relatively
homogeneous class of objects that can be
characterized, on average, by a fixed fraction of the
luminosity of the main galaxy and by typical thickness
and surface brightness values.

\subsection{The length of tidal tails in angular measure}

It is difficult to determine the shape of the tidal tails
that often have an intricate pattern in projection. As
the first approximation, we assumed the tails to be
arcs of circumferences visible at arbitrary angles to
the line of sight. In this case, the observed distribution
of apparent flattenings of the arcs, characterized by the
parameter $k$ that we introduced above, will depend on
the angular measure of the arcs.

Consider the simplest model. Let us take a circumference
of unit radius oriented arbitrarily in space.
Let us choose an arbitrary arc of fixed angular length
$\Delta \varphi$. In projection onto the plane of the sky,
the circumference will appear as an ellipse with an axial
ratio $b/a = \cos i$, where $i$ is the inclination of the
circumference to the plane of the sky. The arc chosen
on the circumference in projection will transform into
an ellipse arc. For the projection of the circumference
arc onto the plane of the sky, we can find
the distance between the extreme points of the arc
$D_{1,\mathrm{mod}}$ and the curvature of the ``elliptic''
arc $D_{2,\mathrm{mod}}$ as the distance between the
straight line connecting
the extreme points of the arc and the tangent to the
arc running parallel to the first straight line.

In the plane of the sky, we will introduce a coordinate
system in such a way that the $x$ axis will
coincide with the line of nodes (the line along which
the circumference intersects with the plane of the sky)
and the $z$ axis will coincide with the line of sight. In
the plane in which the circumference lies, we will also
direct the $x^\prime$ axis along the line of nodes. The angle
between the $y$ and $y^\prime$ axes, just as between the $z$ and
$z^\prime$ axes, will then be equal to $i$.

Let us choose an arc on the circumference. We
will denote the coordinates of its ends by
$(x_1^\prime,y_1^\prime) = (\cos \varphi_1, \sin \varphi_1)$ 
and $(x_2^\prime,y_2^\prime) = (\cos (\varphi_1 + \Delta \varphi_1),
\sin (\varphi_1 + \Delta \varphi_1))$, 
where $\varphi_1$ is the angle in the circumference plane measured from
the line of nodes to the initial end of the arc and
$\Delta \varphi$ is the arclenght. The coordinates of the ends
of the arc projected onto the plane of the sky will then be
$(x_1,y_1) = (\cos \varphi_1, \sin \varphi_1 \, \cos i)$ 
and $(x_2,y_2) = (\cos (\varphi_1 + \Delta \varphi_1), \sin (\varphi_1 + \Delta
\varphi_1)  \, \cos i)$. 
From very simple geometrical considerations, we can obtain
the following expression for the distance between the
ends of the projected arc:
\begin{equation}
D_{1,\mathrm{mod}}^2 = 
4\, \sin^2 \frac{\Delta \varphi}{2} 
\left[ 1 - 
\cos^2 \left(\varphi_1 + \frac{\Delta \varphi}{2}\right) \sin^2 i \right].\,
\end{equation}

The curvature of the arc can be expressed from the
same geometrical considerations as
\begin{equation}
D_{2,\mathrm{mod}}^2 = 
\frac{\left(1 - \sin 
\displaystyle \frac{\Delta \varphi}{2}\right)^2 \, \cos^2 i}
{1 - \cos^2 \left(\varphi_1 + 
\displaystyle \frac{\Delta \varphi}{2}\right) \sin^2 i}.\,
\end{equation}

We will write the parameter characterizing the
flattening of the projected arc, 
$k_\mathrm{mod} = D_{2,\mathrm{mod}} / D_{1,\mathrm{mod}}$,
as
\begin{equation}
k_\mathrm{mod} = \frac{1}{2} \mathrm{tg} \left( \frac{\Delta \varphi}{4} \right)
\frac{\cos i}
{1 - \cos^2 \left(\varphi_1 + \frac{\Delta \varphi}{2}\right) \sin^2 i}.
\end{equation}

The distribution function of the angle $i$ between
the fixed plane (the plane of the sky) and an arbitrary
plane is $f(i) \, d\,i = \sin i \, d\,i$, where $i \in [0, \pi/2]$
(Agekyan 1974). The angle  $\varphi_1$ randomly runs the values from $0$ 
to $2\, \pi$.

\begin{figure}
\centerline{\psfig{file=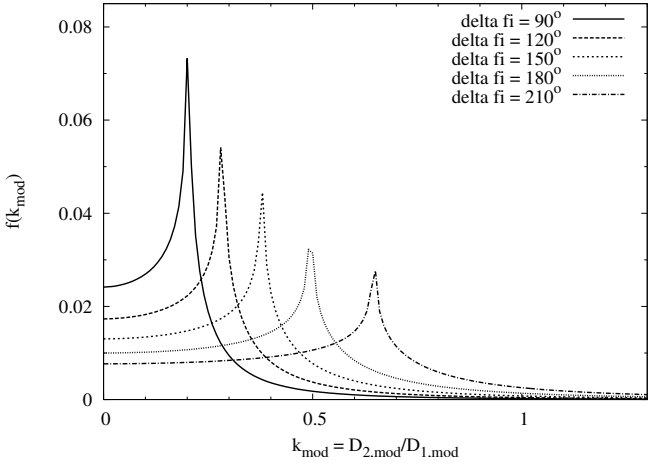,width=9cm,angle=-90,clip=}}
\caption{The distribution of 
$k_\mathrm{mod} = D_{2,\mathrm{mod}} / D_{1,\mathrm{mod}}$,
characterizing
the curvature of the projection of the circumference arc onto
the plane of the sky for various arc lengths in angular measure $\Delta \varphi$.}
\end{figure}

We randomly generated the values of $i$ and $\varphi_1$
in accordance with their distribution functions, determined
$k_\mathrm{mod}$ for each realization at fixed $\Delta \varphi$
and constructed the distribution of $k_\mathrm{mod}$. Figure~6 reproduces
this distribution for several values of $\Delta \varphi$.
The distributions have a maximum near
$$
\displaystyle k_\mathrm{mod}^\mathrm{max} = 
\frac{1}{2} \mathrm{tg} \left( \frac{\Delta \varphi}{4} \right) \, .
$$
Such a curvature of the arc is obtained when $i = 0$. In
this case, any arbitrary segment of the circumference
arc of fixed angular measure $\Delta \varphi$
will give the same value of $k_\mathrm{mod}$, 
which is responsible for the existence of the maximum
$k_\mathrm{mod}^\mathrm{max}$. 

The second prominent feature is the presence of
a long plateau in the range of small $k_\mathrm{mod}$. The arc
is seen as an almost straight line if it is observed at
an angle $i$ close to 90$^{\rm o}$. In this case, the values of
$k_\mathrm{mod}$ will be very low virtually irrespective of the choice
$\varphi_1$ (the initial point of the arc on the circumference)
and the total contribution from these projections to
the distribution function will remain at a fairly high
level, producing an extended plateau.

\begin{figure}
\centerline{\psfig{file=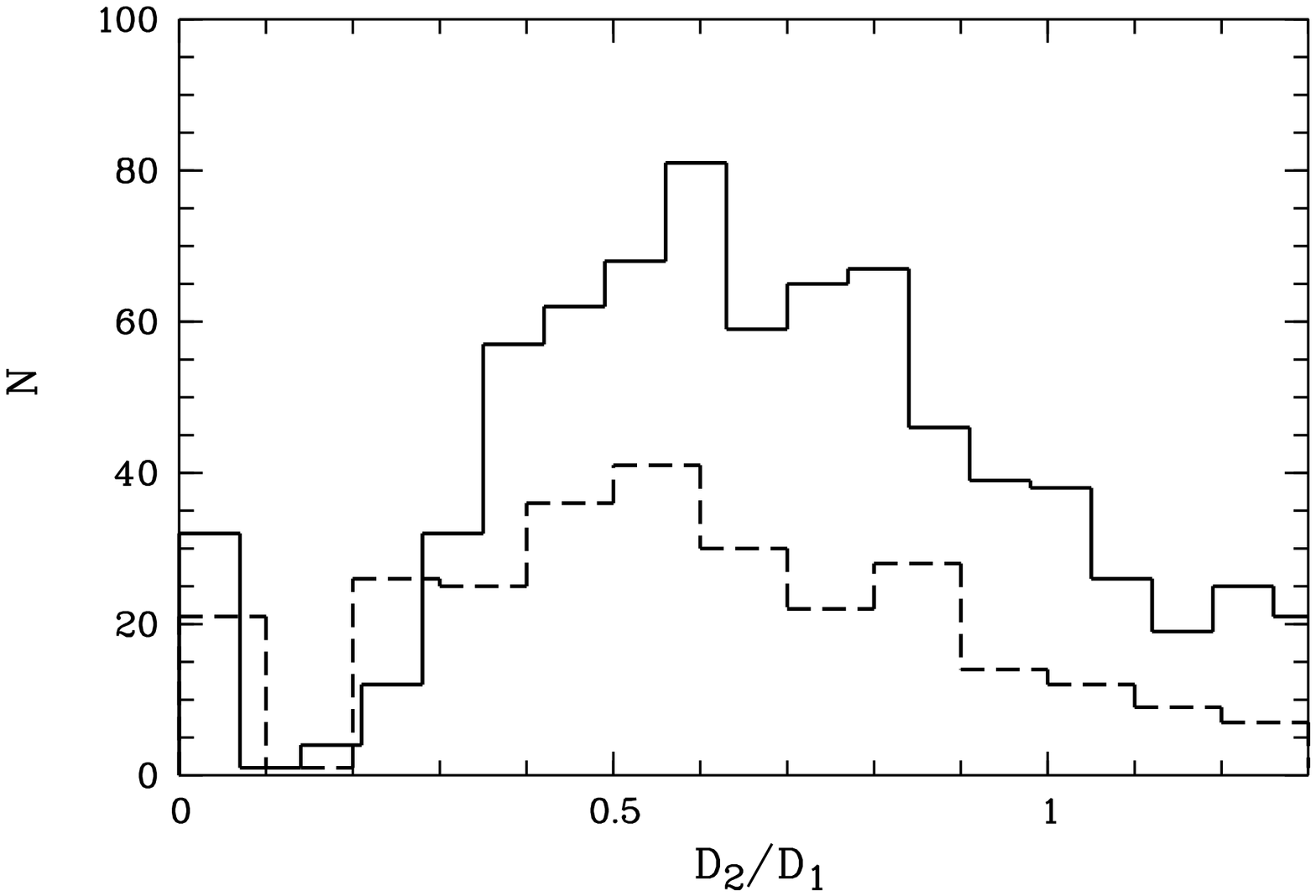,width=8.6cm,angle=-90,clip=}}
\centerline{\psfig{file=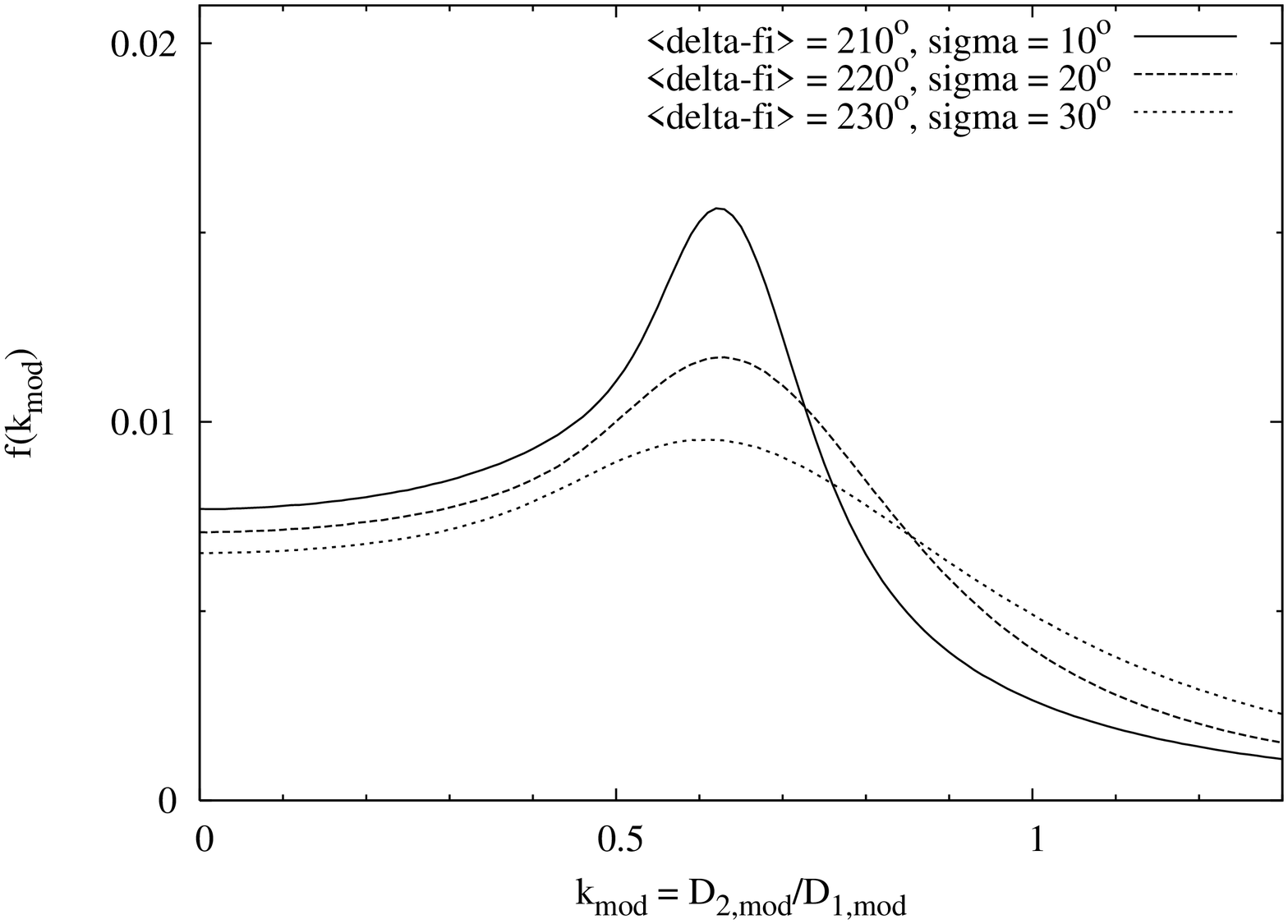,width=9.0cm,angle=-90,clip=}}
\caption{Top: The distributions of nearby (dashed line) and distant (solid line) 
galaxies in $k = D_2/D_1$. Bottom: The model distribution
of $k_\mathrm{mod} = D_{2,\mathrm{mod}} / D_{1,\mathrm{mod}}$  
characterizing the curvature of the projection of the circumference arc onto 
the plane of the sky under the assumption that the arc length in angular 
measure $\Delta \varphi$ is distributed normally.}
\end{figure} 

The observed distribution of $k$ is shown in Fig.~7a.
We see that here there is no such sharp maximum as
that observed for the model distributions. However,
the position of the maximum indicates that the arcs
with angular size $> 180^{\rm o}$ mainly contribute to the
distribution of $k$.

We generated a model distribution of $k_\mathrm{mod}$ for arcs
for which the length  $\Delta \varphi$ was not strictly fixed but had
a scatter $\sigma$ around its mean value ($\Delta \varphi$) 
in accordance with the normal distribution. The model distributions
are shown in Fig.~7b. The data in the figure confirm
our conclusion that, if the tidal tails are assumed to
be arcs of a circumference, then their angular length
is, on average, greater than 180 degrees, i.e., they wind
around the disks of galaxies by more than one half
turn.

\section{Conclusions}

We investigated the geometrical characteristics of
the tidal tails in large samples of nearby and distant
interacting galaxies. It turned out that the visually
distinguished tidal structures, on average, constitute
a homogeneous class of objects that can be characterized
by typical linear and angular lengths, surface
brightness, and total luminosity at a given galaxy
luminosity. Of course, this conclusion refers only to
optically emitting structures, i.e., those containing a
considerable number of stars, that are relatively easily
discernible in digital surveys.

The observed tail length for distant galaxies turned
out to be, on average, smaller than that for nearby
ones. This is probably due to the combination of selection 
effects and the
actual evolution of galaxy properties. The cosmological
dimming and the influence of the $k$ correction can
be attributed to the selection effects, as a result of which 
we observe only the brightest regions of tidal
structures in distant objects. In addition, as a possible
result of the decline in brightness with $z$, in distant
galaxies we predominantly observe a relatively early
evolutionary stage of the tails ($\sim10^8$ yrs
at $z=1$, see Mihos 1995), when they are still relatively 
bright and short, while in nearby objects we see, on average,
``older'' ($\sim10^9$ yrs) and longer structures.

On the other hand, present-day observations and
models of the evolution of galaxies show that compared
to their current characteristics, the spiral galaxies
at $z\sim0.5-1$ must have been a factor 1.5--2
smaller in size and less massive by approximately the
same factor (see, e.g., Dutton et al. 2011). This may
imply that at the evolutionary phase of tidal structures
when we predominantly recognize them by visually
analyzing the images of galaxies, i.e., when their
linear length is comparable to the size of the main
galaxy (see Fig.~3) and the angular length reaches
$\geq180^{\rm o}$ (Fig.~7), they will be, on average, shorter than
those in nearby objects.

The reasons noted above make it much more difficult
to use the statistics of the lengths of tidal tails
for galaxies at various $z$ to study the properties of
their dark halos. Only a detailed simulation of specific
interacting systems can probably help establish well-defined
relationships between the tail lengths and the
dynamical properties of galaxies at various redshifts.

\bigskip
\section*{Acknowledgments}
This work was supported by the Russian Foundation
for Basic Research (project no. 11-02-00471).

\section*{REFERENCES}

\indent

1. T.A. Agekyan, Probability Theory for Astronomers
and Physicists (Nauka, Moscow, 1974) [in Russian].

2. I. Balestra, V. Mainieri, P. Popesso, et al., Astron.
Astrophys. 512, 12 (2010).

3. J. Barnes and L. Hernquist, Nature 360, 715 (1992).

4. C.R. Bridge, R.G. Carlberg, and M. Sullivan, Astrophys.
J. 709, 1067 (2010).

5. D. Coe, N. Benitez, S.F. Sanchez, et al., Astron. J.
132, 926 (2006).

6. J. Dubinski, J.Ch. Mihos, and L. Hernquist, Astrophys.
J. 462, 576 (1996).

7. J. Dubinski, J.Ch. Mihos, and L. Hernquist, Astrophys.
J. 526, 607 (1999).

8. P.-A. Duc, arXiv:1101.4834v2 (2011).

9. A.A. Dutton, F.C. van den Bosch, S.M. Faber, et al.,
Mon. Not. R. Astron. Soc. 410, 1660 (2011).

10. B.G. Elmegreen, M. Kaufman, and M. Thomasson,
Astrophys. J. 412, 90 (1993).

11. D.M. Elmegreen, B.G. Elmegreen, Th. Ferguson,
and B. Mullan, Astrophys. J. 663, 734 (2007).

12. A. Fernandez-Soto, K.M. Lanzetta, and A. Yahil,
Astrophys. J. 513, 34 (1999).

13. K. Glazebrook, A. Verma, B. Boyle, et al., Astron.
J. 131, 2383 (2006).

14. I.D. Karachentsev, Binary Galaxies (Nauka,
Moscow, 1987) [in Russian].

15. J.Ch. Mihos, Astrophys. J. 438, L75 (1995).

16. J.Ch. Mihos, J. Dubinski, and L. Hernquist, Astrophys.
J. 494, 183 (1998).

17. Y.H. Mohamed and V.P. Reshetnikov, Astrofizika
54, 181 (2011).

18. P.B. Nair and R.G. Abraham, Astrophys. J. Suppl.
Ser. 186, 427 (2010).

19. V.P. Reshetnikov, Astron. Lett. 24, 153 (1998).

20. V.P. Reshetnikov, Astron. Astrophys. 353, 92 (2000).

21. V.P. Reshetnikov and N.Ya. Sotnikova, Astron. Astrophys.
Trans. 20, 111 (2001).

22. M. Sawicki and G. Mallen-Ornelas, Astron. J. 126,
1208 (2003).

23. J.M. Schombert, J.F. Wallin, and C. Struck-Marcell,
Astron. J. 99, 497 (1990).

24. N.Ya. Sotnikova and V.P. Reshetnikov, Izv. RAN 62, 1757 (1998a).

25. N.Ya. Sotnikova and V.P. Reshetnikov, Astron. Lett.
24, 73 (1998b).

26. V. Springel and S.D.M. White, Mon. Not.R. Astron.
Soc. 307, 162 (1999).

27. A. Toomre and J. Toomre, Astrophys. J. 178, 623
(1972).

28. R.E. Williams, S. Baum, L.E. Bergeron, et al.,
Astron. J. 120, 2735 (2000).

29. C. Wolf, K. Meisenheimer, M. Kleinheinrich, et al.,
Astron. Astrophys. 421, 913 (2004).

\end{document}